\documentclass[journal=nalefd,manuscript=letter]{achemso}

\pdfoutput=1

\usepackage[active]{srcltx}
\usepackage{graphicx}
\usepackage{mathcomp}	
\usepackage{hyperref}
\usepackage{color}
\usepackage{amsmath,amssymb,amstext,graphicx,bm}
\usepackage{braket}

\author{Hannes~Watzinger}
\email{hannes.watzinger@ist.ac.at}
\affiliation[IST Austria]{Institute of Science and Technology Austria, Am Campus 1, 3400 Klosterneuburg, Austria}

\author{Josip~Kuku\v{c}ka}
\email{josip.kukucka@ist.ac.at}
\affiliation[IST Austria]{Institute of Science and Technology Austria, Am Campus 1, 3400 Klosterneuburg, Austria}

\author{Lada~Vuku\v{s}i\'c}
\affiliation[IST Austria]{Institute of Science and Technology Austria, Am Campus 1, 3400 Klosterneuburg, Austria}

\author{Fei~Gao}
\affiliation{National Laboratory for Condensed Matter Physics and Institute of Physics, Chinese Academy of Sciences, Beijing 100190, China}

\author{Ting~Wang}
\affiliation{National Laboratory for Condensed Matter Physics and Institute of Physics, Chinese Academy of Sciences, Beijing 100190, China}

\author{Friedrich~Sch\"affler}
\affiliation[JK University]{Johannes Kepler University, Institute of Semiconductor and Solid State Physics, Altenbergerstr. 69, 4040 Linz, Austria}

\author{Jian-Jun~Zhang}
\affiliation{National Laboratory for Condensed Matter Physics and Institute of Physics, Chinese Academy of Sciences, Beijing 100190, China}

\author{Georgios~Katsaros}
\affiliation[IST Austria]{Institute of Science and Technology Austria, Am Campus 1, 3400 Klosterneuburg, Austria}

\title{Ge hole spin qubit}

\begin{document}

\begin{abstract}

Holes confined in quantum dots have gained considerable interest in the past few years due to their potential as spin qubits. Here we demonstrate double quantum dot devices in Ge hut wires. Low temperature transport measurements reveal Pauli spin blockade. We demonstrate electric-dipole spin resonance by applying a radio frequency electric field to one of the electrodes defining the double quantum dot. Next, we induce coherent hole spin oscillations by varying the duration of the microwave burst. Rabi oscillations with frequencies reaching 140\,MHz are observed. Finally, Ramsey experiments reveal dephasing times of 130\,ns. The reported results emphasize the potential of Ge as a platform for fast and scalable hole spin qubit devices.

\end{abstract}

\newpage

\begin{figure*}
	\center
  \includegraphics[scale = 1]{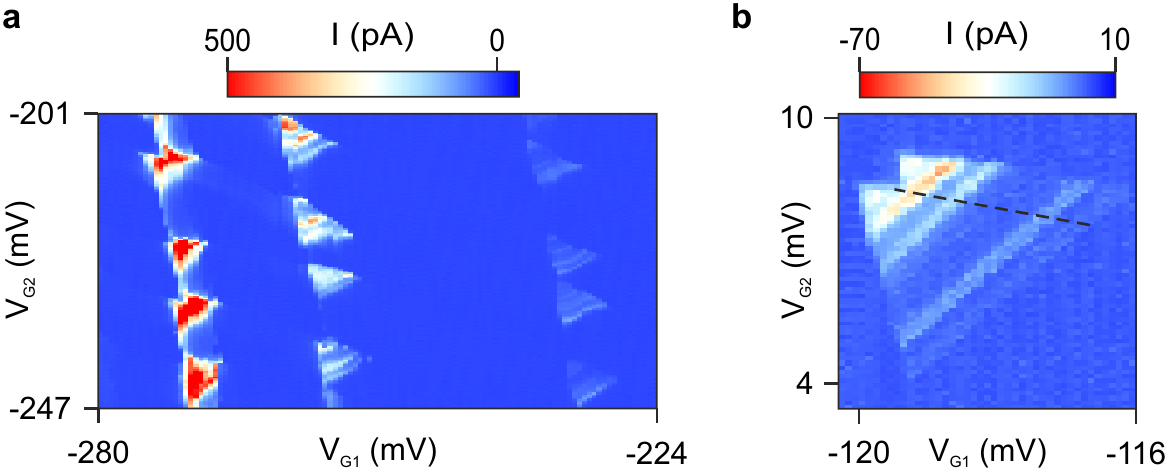}
	\caption{ \textbf{DQD stability diagram from HWs} (a) Stability diagram of a DQD device showing the characteristic bias triangles at a bias voltage of $V_{\textrm{SD}} = 2\,mV$. (b) Representative zoom-in of a pair of bias triangles at $V_{\textrm{SD}} = -2\,mV$ from a second device. The dashed black line indicates the edge of the lower bias triangle.} 
	\label{fig::ImageI}
\end{figure*}

\begin{figure*}
	\center
  \includegraphics[scale = 1]{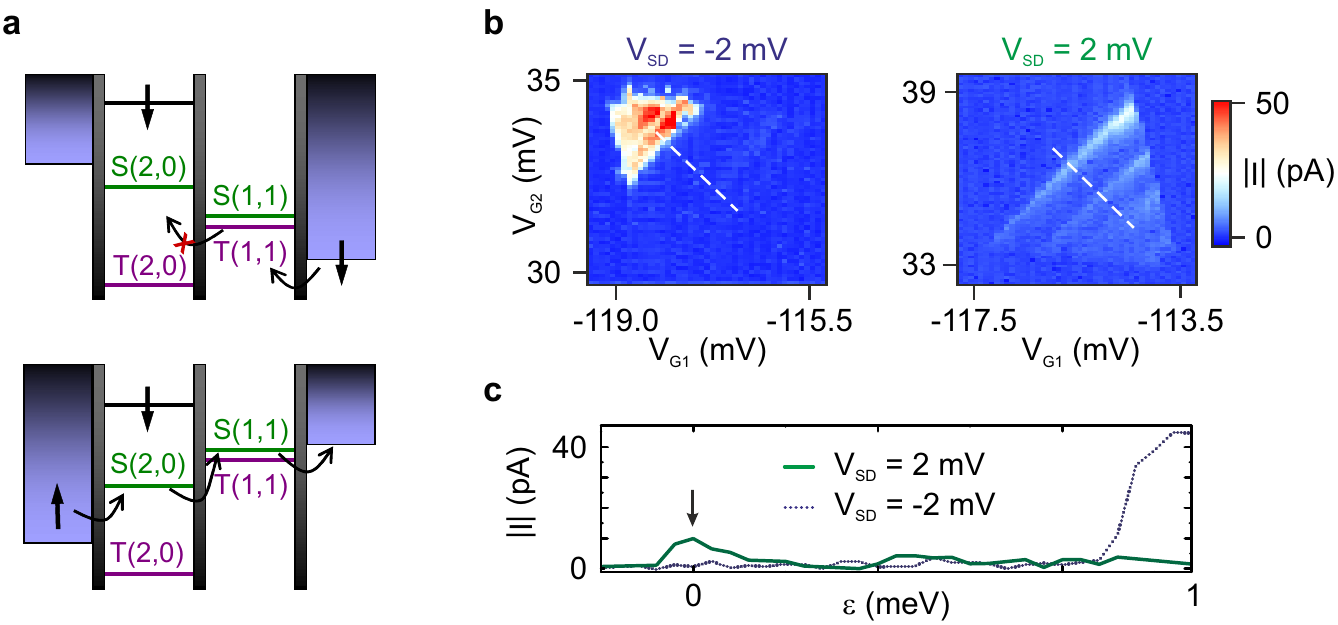}
	\caption{\textbf{Spin blockade in DQDs} (a) Schematic presentation of PSB for a hole DQD. Transport is blocked for the transition (1,1) $\rightarrow$ (0,2) (upper panel) due to the Pauli exclusion principle and can be lifted by reversing the applied bias voltage (lower panel). (b) Bias triangles exhibiting PSB at negative bias voltages (left). Reversing the bias results in an enhancement of the baseline current (right). (c) Comparison of the current from two line cuts along the detuning axis for positive (green solid line) and negative (blue dotted line) bias voltages. The positions where the line cuts were taken are indicated by white dashed lines in (b).} 
	\label{fig::ImageII}
\end{figure*}

\begin{figure*}
	\center
  \includegraphics[scale = 1]{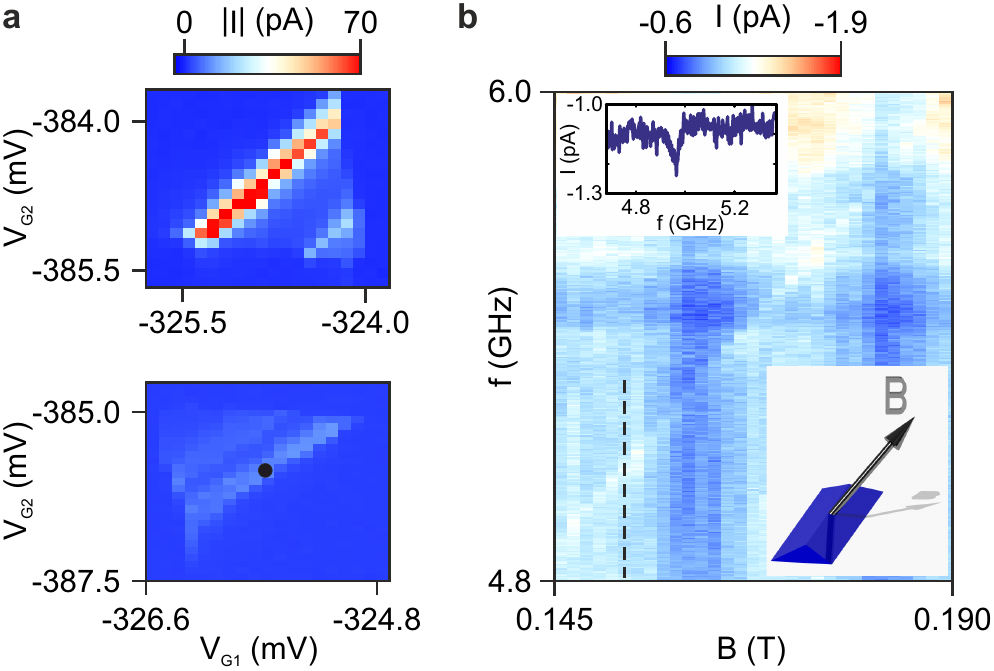}
	\caption{\textbf{EDSR Spectroscopy} (a) Set of bias triangles from a third device for $V_{\textrm{SD}} = 1\,mV$ (upper panel) and $V_{\textrm{SD}} = -1\,mV$ (lower panel). (b) Raw data measurement showing the frequency versus magnetic field dependence of the zero detuning current measured at the position marked by the black circle in (a). The magnetic field is oriented 45\,$^{\circ}$ in respect to both the out-of-plane direction and the HW axes (see lower inset). A magnetic field offset of about 30\,mT is caused by the hysteretic behavior of the magnet. The upper inset shows a line trace taken along the black dashed line.} 
	\label{fig::ImageIII}
\end{figure*}

\begin{figure*}
	\center
  \includegraphics[scale = 1]{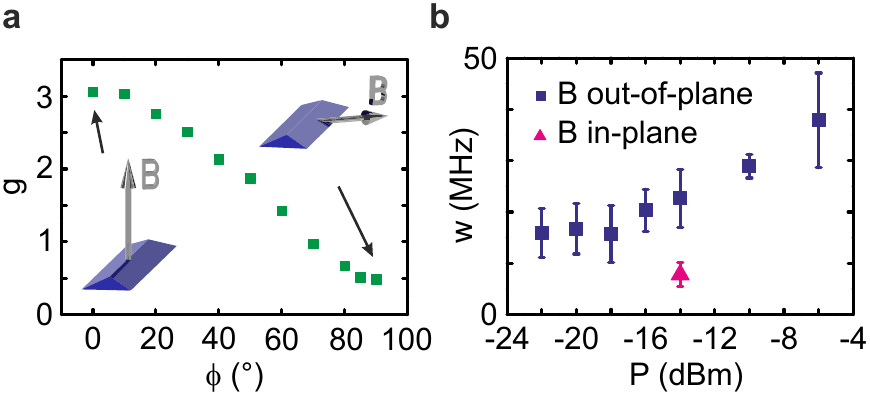}
	\caption{\textbf{Angle dependence of the g-factor and power dependence of the EDSR peak width} (a) Extracted g-factor values for different magnetic field orientations as illustrated by the two insets. $\phi$ denotes the angle of the magnetic field vector $\vec{B}$ to the [001] direction, i.e. $\phi$ = $0^{\circ}$ ($90^{\circ}$) corresponds to an out-of-plane (in-plane) magnetic field. The g-factors were extracted from a linear fit to the values obtained by a Gaussian fit to the EDSR peaks at several positions. The errors are below 0.03 for all data points and therefore not visible. The measurements were taken with a RF power of -14\,dBm. (b) Power dependence of the FWHM ($w$) of the resonance peak from Fig. \ref{fig::ImageIII}b for an out-of-plane magnetic field (blue squares). Below -16\,dBm the width saturates at around 16\,MHz. However, for an in-plane magnetic field and for a power of -14\,dBm the EDSR peak width shrinks to $7.8 \pm 2.3$ MHz (red triangle). For lower power values the EDSR line is not any more visible. The error bars were extracted from averaging over all values obtained from Gaussian fits to the resonance peaks at the respective power.} 
	\label{fig::ImageIV}
\end{figure*}

\begin{figure}
	\center
  \includegraphics[scale = 1]{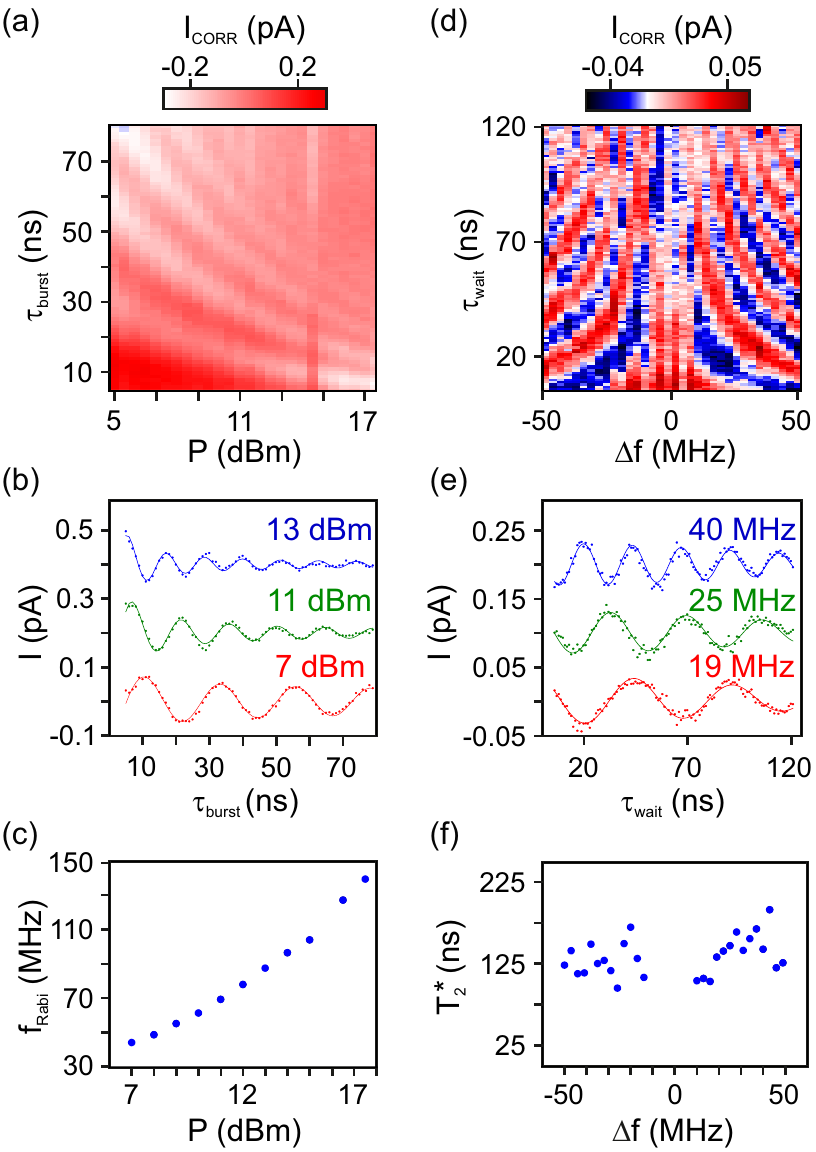}
	\caption{\textbf{Rabi oscillations and Ramsey fringes} (a) Current versus microwave burst time $\tau_\textrm{burst}$ and power for an out-of-plane magnetic field of 127\,mT and an RF frequency of 5.96555\,GHz for the third device after thermal cycling. Rabi oscillations can be observed. The data has been corrected by removing the average column and row values from the corresponding data points. (b) Characteristic traces showing Rabi oscillations versus $\tau_\textrm{burst}$. The traces  for 11 and 13\,dBm have been shifted by 200 and 400\,fA, respectively. For extracting the Rabi frequencies the uncorrected data were fitted to $\frac{A}{\tau_\textrm{burst}^a} sin(2\pi f_\textrm{Rabi} \tau_\textrm{burst} + \phi) + \textrm{offset} + c \sqrt{\tau_\textrm{burst}}$. The offset and $c \sqrt{\tau_\textrm{burst}}$ have been subtracted from the shown traces. (c) Dependence of the Rabi frequency versus the applied RF power. (d) Current versus frequency and waiting time for a power equal to 11\,dBm, an out-of-plane magnetic field of 127\,mT and a center frequency of 5.96555\,GHz. (e) Traces of the current versus $\tau_\textrm{wait}$ for frequency detunings of 19, 25 and 40\,MHz. The traces for 25 and 40\,MHz are shifted by 100 and 200\,fA, respectively. (f) Plot showing the extracted dephasing times for various detunings. The times were extracted by fitting the data to $A e^{-\left( \frac{\tau_\textrm{wait}}{T_2^*} \right)^2} sin(2\pi \Delta f \tau_\textrm{wait} + \phi) + \textrm{offset}$. The average dephasing time exceeds 130\,ns.} 
	\label{fig::ImageV}
\end{figure}

Spins in isotopically purified Si have shown record coherence times \cite{Muhonen2014} and fidelities \cite{Yoneda2017} making them promising candidates for scalable quantum circuits \cite{Vandersypen2017}. One of the key ingredients for realizing such circuits will be a strong coupling of spins to superconducting resonators \cite{Viennot2015}. This has been recently achieved for Si by dressing electrons with spin orbit coupling \cite{Mi2017, Samkharadze2017}. Holes, on the other hand, have intrinsically strong spin orbit coupling making them thus promising candidates for electrically controlled and scalable qubits. In 2016, the first hole spin qubit was reported in a Si FinFET \cite{Maurand2016}. Beside Si, also Ge has emerged as an interesting material system due to the strong and tunable spin orbit coupling \cite{Hao2010, Higginbotham2014, Kloeffel2017, Marcellina2017} good contacts to superconductors \cite{Xiang2006, Katsaros2010, Veldhorst2018} and relatively small effective mass. Indeed, in the past few years quantum dot (QD) devices in Ge/Si core/shell nanowires \cite{Hu2012, Higginbotham2014, Brauns:PRB2016, Hu2007}, self-assembled nanocrystals \cite{Katsaros2010} and Ge hole gases \cite{Veldhorst2018} have been reported. However, no qubit has been demonstrated so far.
\newline 
In this letter, we demonstrate the ability to capture holes in double quantum dots (DQD) fabricated in Ge hut wires (HWs) \cite{Zhang2012, Watzinger2014}. We make use of the Pauli spin blockade (PSB) \cite{Ono2002} mechanism and the electric-dipole spin resonance (EDSR) technique in order to demonstrate the addressability of single holes. By varying the duration of the microwave burst, Rabi oscillations with frequencies higher than  100\,MHz have been observed. Finally, Ramsey fringes-like measurements reveal dephasing times of 130\,ns, twice the dephasing time reported for holes in Si \cite{Maurand2016}.
\newline
The stability diagram of a DQD device showing characteristic bias triangles is depicted in Fig. 1a with the two gate voltages $V_\textrm{G1}$ and $V_\textrm{G2}$.
A representative measurement of two bias triangles from a second device is shown in Fig. \ref{fig::ImageI}b. Due to the fairly low mutual capacitance of about 1\,aF they are merged already at relatively low bias voltages. The ground state as well as several additional excited states are clearly visible.
Energy level separations up to 1\,meV  and a relative lever arm $\Delta V_\textrm{G1} / \Delta V_\textrm{G2} = 0.7$ are observed. Since the two top gates G1 and G2 are very close to the HW a relatively strong coupling is obtained, leading to alpha factors of $\alpha_1 = 0.62$ and $\alpha_2 = 0.43$ \cite{VanderWiel2002}. 
\newline
In order to realize a spin 3/2 qubit in our DQD devices we rely on Pauli spin blockade (PSB) \cite{Ono2002} as a spin-selective read-out mechanism \cite{Hanson2007}.
PSB occurs in a (1,1) $\rightarrow$ (0,2) or an equivalent (2N-1, 2N-1) $\rightarrow$ (2N-2, 2N) charge configuration (see Fig. \ref{fig::ImageII}a).
In such a configuration transport through the DQD is blocked due to spin selectivity even if it is energetically allowed. By reversing the applied source-drain bias voltage the spin blockade can be lifted. Signatures of PSB were observed in several bias triangles exhibiting a suppressed leakage current of the triangle baseline.
Two representative direct current (DC) measurements are shown in the left and right panel of Fig. \ref{fig::ImageII}b for bias voltages of -2\,mV and +2\,mV, respectively.
The corresponding line traces along the detuning direction (white dashed lines) are plotted below in Fig. \ref{fig::ImageII}c. In the blocked configuration (blue dotted line) the zero-detuning current, indicated by the black arrow, drops to about 2\,pA compared to 10\,pA in the non-blocked case (green solid line), as expected for PSB \cite{Hanson2007}. 
\newline
Rotating one of the spins can lift PSB. This can be achieved via the EDSR mechanism \cite{Golovach2006}. An alternating current (AC) electric field applied to one of the gates of the DQD (here G1) can cause oscillations in the position of the confined hole wave function (see Fig. \ref{fig::ImageIII}b). Such an oscillation in combination with a constant applied magnetic field can lead to spin rotations in systems with strong spin orbit coupling \cite{Hanson2007}. In order to induce such continuous wave spin rotations the driving frequency of the AC electric field has to be equal to the Larmor frequency $f_0 = \left|g\right|\mu_\textrm{B}B/h$, where $g$ is the g-factor for a certain magnetic field orientation, $\mu_\textrm{B}$ is the Bohr magneton and $h$ is Planck's constant.
Fig. \ref{fig::ImageIII}a shows a pair of bias triangles for positive and negative bias voltages from a third measured device. 
The black circle in the lower panel of Fig. \ref{fig::ImageIII}a indicates the position at which the EDSR measurement shown in Fig. \ref{fig::ImageIII}c was performed. From the slope of the resonance line a g-factor of about 2 can be extracted.  
\newline
By changing the direction of the magnetic field the slope of the EDSR line is changing due to the direction dependence of the g-factor. Each of the g-factor values shown in Fig. \ref{fig::ImageIV}a was extracted from a linear fit through several points along the respective resonance line. The g-factor values show a strong anisotropy in good agreement with earlier experimental findings for HH states\cite{Watzinger2016}.
EDSR does not only lift PSB, but also allows the extraction of a lower limit for the hole spin dephasing time $T_{2}^*$. 
In order to extract a lower bound for the dephasing time $T_{2}^*$ the power of the applied radio frequency (RF) signal was varied. At high power, the EDSR width is power broadened. However, for measurements taken in an out-of-plane magnetic field the width is saturating at values of about -18\,dBm, as can be seen in Fig. \ref{fig::ImageIV}b.  
Therefore, a lower bound for the dephasing time of about 33\,ns can be extracted using the relation $T_{2}^* = 2\sqrt{ln(2)}/\pi w$, where $w$ is the full width at half maximum (FWHM) of the resonance peak at a certain RF power\cite{Kawakami2014}. 
For HH states it has been predicted that the direction of the applied magnetic field has a strong influence on the dephasing times \cite{Fischer2008}. Indeed, optical measurements of hole spins confined in GaAs self-assembled QDs have shown very long dephasing times \cite{Prechtel2016}. In order to obtain such longer dephasing times, the external magnetic field needs to be aligned perpendicular to the direction of the Overhauser field, which for HH states is perpendicular to the growth plane \cite{Fischer2008}. By repeating the EDSR measurement for an in-plane magnetic field and a RF power of -14\,dBm, we obtain a 68\,ns dephasing time (see Supplementary Information). 
\newline
In order to demonstrate coherent control over the hole spin state, periodic square pulses are applied to G1 in order to shift between the spin blockade and Coulomb blockade configuration. The modulation period is fixed to 310\,ns out of which 100\,ns were devoted for spin manipulation and the rest for the read-out and the initialization. The system is initialized in the triplet state. When in Coulomb blockade, a microwave burst of varying duration is applied. For a $\pi$-pulse the hole spin will flip leading thus to a singlet (1,1) state. The system is then brought back into the spin blockade region for spin read-out and the hole can tunnel to the singlet (2,0) state leading to an enhanced current. By varying the duration of the microwave burst, oscillations of the detected current can be observed [Fig. \ref{fig::ImageV}(a)].
As expected, the period of the Rabi oscillations decreases with increasing power of the microwave burst [see Fig. \ref{fig::ImageV}(b)].  Rabi frequencies approaching 140\,MHz are observed [Fig. \ref{fig::ImageV}(c)]. They are faster than what has been predicted for Ge nanocrystals \cite{Ares2013} and than those reported for the InSb electron spin qubit which showed 8\,ns dephasing time \cite{Berg2013}. 
To measure the inhomogeneous depashing time $T_2^*$, Ramsey experiments were performed. For this purpose, a pulse sequence consisting of two short $\frac{\pi}{2}$-pulses separated by a waiting time $\tau_\textrm{wait}$, during which the qubit can freely evolve and dephase, was applied. For each frequency the current oscillates as a function of $\tau_\textrm{wait}$ [Fig. \ref{fig::ImageV}(d)]. From the decay time of these oscillations, average dephasing times exceeding 130\,ns have been measured [Figs. \ref{fig::ImageV}(e) and (f)]. Due to the limited visibility in our experiment, caused by the small current flowing through the DQD, it was not possible to extend the waiting time further than 160\,ns. This prohibited the investigation of longer $T_2^*$ possibly arising for parallel magnetic fields as has been shown in Fig. \ref{fig::ImageIV}(b). 
\newline
In conclusion, by using PSB in a DQD device we have demonstrated the first Ge hole spin qubit with Rabi frequencies reaching 140\,MHz. Despite the strong spin orbit coupling, the obtained $T_2^*$ is higher than that of holes \cite{Maurand2016} confined in QDs formed in natural Si and less than one order of magnitude lower than that of electrons \cite{Kawakami2014}. The reported result paves the way towards long-range two qubit-gates \cite{Nigg2017,Kloeffel:PRB2013}. 
\newline
We thank A. Hofmann, E. Laird, R. Maurand, J. Petta and M. Veldhorst for helpful discussions. This research was supported by the Scientific Service Units of IST Austria through resources provided by the MIBA Machine Shop and the nanofabrication facility. We acknowledge financial support by the Austrian Ministry of Science through the HRSM call 2016. The work was also supported by the ERC Starting Grant no. 335497, the FWF-Y 715-N30 project, the National Key R\&D Program of China (Grant No. 2016YFA0301701) and the NSFC (Grants No. 11574356 and 11434010).
\newline
H.W. and J.K. contributed equally to this work.

\newpage

\bibliography{bibliography}

\providecommand{\latin}[1]{#1}
\makeatletter
\providecommand{\doi}
  {\begingroup\let\do\@makeother\dospecials
  \catcode`\{=1 \catcode`\}=2\doi@aux}
\providecommand{\doi@aux}[1]{\endgroup\texttt{#1}}
\makeatother
\providecommand*\mcitethebibliography{\thebibliography}
\csname @ifundefined\endcsname{endmcitethebibliography}
  {\let\endmcitethebibliography\endthebibliography}{}
\begin{mcitethebibliography}{32}
\providecommand*\natexlab[1]{#1}
\providecommand*\mciteSetBstSublistMode[1]{}
\providecommand*\mciteSetBstMaxWidthForm[2]{}
\providecommand*\mciteBstWouldAddEndPuncttrue
  {\def\EndOfBibitem{\unskip.}}
\providecommand*\mciteBstWouldAddEndPunctfalse
  {\let\EndOfBibitem\relax}
\providecommand*\mciteSetBstMidEndSepPunct[3]{}
\providecommand*\mciteSetBstSublistLabelBeginEnd[3]{}
\providecommand*\EndOfBibitem{}
\mciteSetBstSublistMode{f}
\mciteSetBstMaxWidthForm{subitem}{(\alph{mcitesubitemcount})}
\mciteSetBstSublistLabelBeginEnd
  {\mcitemaxwidthsubitemform\space}
  {\relax}
  {\relax}

\bibitem[Muhonen \latin{et~al.}(2014)Muhonen, Dehollain, Laucht, Hudson, Kalra,
  Sekiguchi, Itoh, Jamieson, McCallum, Dzurak, and Morello]{Muhonen2014}
Muhonen,~J.~T.; Dehollain,~J.~P.; Laucht,~A.; Hudson,~F.~E.; Kalra,~R.;
  Sekiguchi,~T.; Itoh,~K.~M.; Jamieson,~D.~N.; McCallum,~J.~C.; Dzurak,~A.~S.;
  Morello,~A. \emph{Nature Nanotechnology} \textbf{2014}, \emph{9},
  986--991\relax
\mciteBstWouldAddEndPuncttrue
\mciteSetBstMidEndSepPunct{\mcitedefaultmidpunct}
{\mcitedefaultendpunct}{\mcitedefaultseppunct}\relax
\EndOfBibitem
\bibitem[Jun~Yoneda \latin{et~al.}(2017)Jun~Yoneda, Takeda, Otsuka, Nakajima,
  Delbecq, Allison, Honda, Kodera, Oda, Hoshi, Usami, Itoh, and
  Tarucha]{Yoneda2017}
Jun~Yoneda,~J.; Takeda,~K.; Otsuka,~T.; Nakajima,~T.; Delbecq,~M.~R.;
  Allison,~G.; Honda,~T.; Kodera,~T.; Oda,~S.; Hoshi,~Y.; Usami,~N.;
  Itoh,~K.~M.; Tarucha,~S. \emph{Nature Nanotechnology} \textbf{2017}, \relax
\mciteBstWouldAddEndPunctfalse
\mciteSetBstMidEndSepPunct{\mcitedefaultmidpunct}
{}{\mcitedefaultseppunct}\relax
\EndOfBibitem
\bibitem[Vandersypen \latin{et~al.}(2017)Vandersypen, Bluhm, Clarke, Dzurak,
  Ishihara, Morello, Reilly, R., and Veldhorst]{Vandersypen2017}
Vandersypen,~L. M.~K.; Bluhm,~H.; Clarke,~J.~S.; Dzurak,~A.~S.; Ishihara,~R.;
  Morello,~A.; Reilly,~D.~J.; R.,~S.~L.; Veldhorst,~M. \emph{npj Quantum
  Information} \textbf{2017}, \emph{3}\relax
\mciteBstWouldAddEndPuncttrue
\mciteSetBstMidEndSepPunct{\mcitedefaultmidpunct}
{\mcitedefaultendpunct}{\mcitedefaultseppunct}\relax
\EndOfBibitem
\bibitem[Viennot \latin{et~al.}(2015)Viennot, Dartiailh, Cottet, and
  Kontos]{Viennot2015}
Viennot,~J.~J.; Dartiailh,~M.~C.; Cottet,~A.; Kontos,~T. \emph{Science}
  \textbf{2015}, \emph{349}, 408--411\relax
\mciteBstWouldAddEndPuncttrue
\mciteSetBstMidEndSepPunct{\mcitedefaultmidpunct}
{\mcitedefaultendpunct}{\mcitedefaultseppunct}\relax
\EndOfBibitem
\bibitem[Mi \latin{et~al.}(2017)Mi, Benito, Putz, Zajac, Taylor, Burkard, and
  Petta]{Mi2017}
Mi,~X.; Benito,~M.; Putz,~S.; Zajac,~D.~M.; Taylor,~J.~M.; Burkard,~G.;
  Petta,~J.~R. \emph{arXiv:1710.03265} \textbf{2017}, \relax
\mciteBstWouldAddEndPunctfalse
\mciteSetBstMidEndSepPunct{\mcitedefaultmidpunct}
{}{\mcitedefaultseppunct}\relax
\EndOfBibitem
\bibitem[Samkharadze \latin{et~al.}(2018)Samkharadze, Zheng, Kalhor, Brousse,
  Sammak, Mendes, Blais, Scappucci, and Vandersypen]{Samkharadze2017}
Samkharadze,~N.; Zheng,~G.; Kalhor,~N.; Brousse,~D.; Sammak,~A.; Mendes,~U.~C.;
  Blais,~A.; Scappucci,~G.; Vandersypen,~L. M.~K. \emph{Science} \textbf{2018},
  \relax
\mciteBstWouldAddEndPunctfalse
\mciteSetBstMidEndSepPunct{\mcitedefaultmidpunct}
{}{\mcitedefaultseppunct}\relax
\EndOfBibitem
\bibitem[R.~Maurand \latin{et~al.}(2016)R.~Maurand, Jehl, Kotekar-Patil, Corna,
  Bohuslavskyi, Lavi\'eville, Hutin, Barraud, Vinet, Sanquer, and
  De~Franceschi]{Maurand2016}
R.~Maurand,~R.; Jehl,~X.; Kotekar-Patil,~D.; Corna,~A.; Bohuslavskyi,~H.;
  Lavi\'eville,~R.; Hutin,~L.; Barraud,~S.; Vinet,~M.; Sanquer,~M.;
  De~Franceschi,~S. \emph{Nature Comm.} \textbf{2016}, \emph{7}, 13575\relax
\mciteBstWouldAddEndPuncttrue
\mciteSetBstMidEndSepPunct{\mcitedefaultmidpunct}
{\mcitedefaultendpunct}{\mcitedefaultseppunct}\relax
\EndOfBibitem
\bibitem[Hao \latin{et~al.}(2010)Hao, Tu, Cao, Zhou, Li, Guo, Fung, Ji, Guo,
  and Lu]{Hao2010}
Hao,~X.-J.; Tu,~T.; Cao,~G.; Zhou,~C.; Li,~H.-O.; Guo,~G.-C.; Fung,~W.~Y.;
  Ji,~Z.; Guo,~G.-P.; Lu,~W. \emph{Nano Lett.} \textbf{2010}, \emph{10},
  2956--2960\relax
\mciteBstWouldAddEndPuncttrue
\mciteSetBstMidEndSepPunct{\mcitedefaultmidpunct}
{\mcitedefaultendpunct}{\mcitedefaultseppunct}\relax
\EndOfBibitem
\bibitem[Higginbotham \latin{et~al.}(2014)Higginbotham, Larsen, Yao, Yan,
  Lieber, Marcus, and Kuemmeth]{Higginbotham2014}
Higginbotham,~A.~P.; Larsen,~T.~W.; Yao,~J.; Yan,~H.; Lieber,~C.~M.;
  Marcus,~C.~M.; Kuemmeth,~F. \emph{Nano Lett.} \textbf{2014}, \emph{14},
  3582--3586\relax
\mciteBstWouldAddEndPuncttrue
\mciteSetBstMidEndSepPunct{\mcitedefaultmidpunct}
{\mcitedefaultendpunct}{\mcitedefaultseppunct}\relax
\EndOfBibitem
\bibitem[Kloeffel \latin{et~al.}(2017)Kloeffel, Rančić, and
  Loss]{Kloeffel2017}
Kloeffel,~C.; Rančić,~M.~J.; Loss,~D. \emph{arXiv:1712.03476} \textbf{2017},
  \relax
\mciteBstWouldAddEndPunctfalse
\mciteSetBstMidEndSepPunct{\mcitedefaultmidpunct}
{}{\mcitedefaultseppunct}\relax
\EndOfBibitem
\bibitem[Marcellina \latin{et~al.}(2017)Marcellina, Hamilton, Winkler, and
  Culcer]{Marcellina2017}
Marcellina,~E.; Hamilton,~A.~R.; Winkler,~R.; Culcer,~D. \emph{Phys. Rev. B}
  \textbf{2017}, \emph{95}, 075305\relax
\mciteBstWouldAddEndPuncttrue
\mciteSetBstMidEndSepPunct{\mcitedefaultmidpunct}
{\mcitedefaultendpunct}{\mcitedefaultseppunct}\relax
\EndOfBibitem
\bibitem[Xiang \latin{et~al.}(2006)Xiang, Vidan, Tinkham, Westervelt, and
  Lieber]{Xiang2006}
Xiang,~J.; Vidan,~A.; Tinkham,~M.; Westervelt,~R.~M.; Lieber,~C.~M.
  \emph{Nature Nanotechnology} \textbf{2006}, \emph{1}, 208--213\relax
\mciteBstWouldAddEndPuncttrue
\mciteSetBstMidEndSepPunct{\mcitedefaultmidpunct}
{\mcitedefaultendpunct}{\mcitedefaultseppunct}\relax
\EndOfBibitem
\bibitem[Katsaros \latin{et~al.}(2010)Katsaros, Spathis, Stoffel, Fournel,
  Mongillo, Bouchiat, Lefloch, Rastelli, Schmidt, and
  De~Franceschi]{Katsaros2010}
Katsaros,~G.; Spathis,~P.; Stoffel,~M.; Fournel,~F.; Mongillo,~M.;
  Bouchiat,~V.; Lefloch,~F.; Rastelli,~A.; Schmidt,~O.~G.; De~Franceschi,~S.
  \emph{Nature Nanotechnology} \textbf{2010}, \emph{5}, 458--464\relax
\mciteBstWouldAddEndPuncttrue
\mciteSetBstMidEndSepPunct{\mcitedefaultmidpunct}
{\mcitedefaultendpunct}{\mcitedefaultseppunct}\relax
\EndOfBibitem
\bibitem[N.~W.~Hendrickx \latin{et~al.}(2018)N.~W.~Hendrickx, Franke, Sammak,
  Kouwenhoven, Sabbagh, Yeoh, Li, Tagliaferri, Virgilio, Capellini, Scappucci,
  and Veldhorst]{Veldhorst2018}
N.~W.~Hendrickx,~N.~W.; Franke,~D.~P.; Sammak,~A.; Kouwenhoven,~M.;
  Sabbagh,~D.; Yeoh,~L.; Li,~R.; Tagliaferri,~M.; Virgilio,~M.; Capellini,~G.;
  Scappucci,~G.; Veldhorst,~M. \emph{arXiv:1801.08869} \textbf{2018}, \relax
\mciteBstWouldAddEndPunctfalse
\mciteSetBstMidEndSepPunct{\mcitedefaultmidpunct}
{}{\mcitedefaultseppunct}\relax
\EndOfBibitem
\bibitem[Hu \latin{et~al.}(2012)Hu, Kuemmeth, Lieber, and Marcus]{Hu2012}
Hu,~Y.~J.; Kuemmeth,~F.; Lieber,~C.~M.; Marcus,~C.~M. \emph{Nature
  Nanotechnology} \textbf{2012}, \emph{7}, 47--50\relax
\mciteBstWouldAddEndPuncttrue
\mciteSetBstMidEndSepPunct{\mcitedefaultmidpunct}
{\mcitedefaultendpunct}{\mcitedefaultseppunct}\relax
\EndOfBibitem
\bibitem[Brauns \latin{et~al.}(2016)Brauns, Ridderbos, Li, Bakkers, and
  Zwanenburg]{Brauns:PRB2016}
Brauns,~M.; Ridderbos,~J.; Li,~A.; Bakkers,~E. P. A.~M.; Zwanenburg,~F.~A.
  \emph{Phys. Rev. B} \textbf{2016}, \emph{93}, 121408(R)\relax
\mciteBstWouldAddEndPuncttrue
\mciteSetBstMidEndSepPunct{\mcitedefaultmidpunct}
{\mcitedefaultendpunct}{\mcitedefaultseppunct}\relax
\EndOfBibitem
\bibitem[Hu \latin{et~al.}(2007)Hu, Churchill, Reilly, Xiang, Lieber, and
  Marcus]{Hu2007}
Hu,~Y.~J.; Churchill,~H. O.~H.; Reilly,~D.~J.; Xiang,~J.; Lieber,~C.~M.;
  Marcus,~C.~M. \emph{Nature Nanotechnology} \textbf{2007}, \emph{2},
  622--625\relax
\mciteBstWouldAddEndPuncttrue
\mciteSetBstMidEndSepPunct{\mcitedefaultmidpunct}
{\mcitedefaultendpunct}{\mcitedefaultseppunct}\relax
\EndOfBibitem
\bibitem[Zhang \latin{et~al.}(2012)Zhang, Katsaros, Montalenti, Scopece,
  Rezaev, Mickel, Rellinghaus, Miglio, De~Franceschi, Rastelli, and
  Schmidt]{Zhang2012}
Zhang,~J.~J.; Katsaros,~G.; Montalenti,~F.; Scopece,~D.; Rezaev,~R.~O.;
  Mickel,~C.; Rellinghaus,~B.; Miglio,~L.; De~Franceschi,~S.; Rastelli,~A.;
  Schmidt,~O.~G. \emph{Phys. Rev. Lett.} \textbf{2012}, \emph{109},
  085502\relax
\mciteBstWouldAddEndPuncttrue
\mciteSetBstMidEndSepPunct{\mcitedefaultmidpunct}
{\mcitedefaultendpunct}{\mcitedefaultseppunct}\relax
\EndOfBibitem
\bibitem[Watzinger \latin{et~al.}(2014)Watzinger, Glaser, Zhang, Daruka, and
  Sch\"affler]{Watzinger2014}
Watzinger,~H.; Glaser,~M.; Zhang,~J.~J.; Daruka,~I.; Sch\"affler,~F. \emph{APL
  Mater.} \textbf{2014}, \emph{2}, 076102\relax
\mciteBstWouldAddEndPuncttrue
\mciteSetBstMidEndSepPunct{\mcitedefaultmidpunct}
{\mcitedefaultendpunct}{\mcitedefaultseppunct}\relax
\EndOfBibitem
\bibitem[Ono \latin{et~al.}(2002)Ono, Austing, Tokura, and Tarucha]{Ono2002}
Ono,~K.; Austing,~D.~G.; Tokura,~Y.; Tarucha,~S. \emph{Science} \textbf{2002},
  \emph{297}, 1313--1317\relax
\mciteBstWouldAddEndPuncttrue
\mciteSetBstMidEndSepPunct{\mcitedefaultmidpunct}
{\mcitedefaultendpunct}{\mcitedefaultseppunct}\relax
\EndOfBibitem
\bibitem[van~der Wiel \latin{et~al.}(2002)van~der Wiel, De~Franceschi,
  Elzerman, Fujisawa, Tarucha, and Kouwenhoven]{VanderWiel2002}
van~der Wiel,~W.~G.; De~Franceschi,~S.; Elzerman,~J.~M.; Fujisawa,~T.;
  Tarucha,~S.; Kouwenhoven,~L.~P. \emph{Rev. Mod. Phys.} \textbf{2002},
  \emph{75}\relax
\mciteBstWouldAddEndPuncttrue
\mciteSetBstMidEndSepPunct{\mcitedefaultmidpunct}
{\mcitedefaultendpunct}{\mcitedefaultseppunct}\relax
\EndOfBibitem
\bibitem[Hanson \latin{et~al.}(2007)Hanson, Kouwenhoven, Petta, Tarucha, and
  Vandersypen]{Hanson2007}
Hanson,~R.; Kouwenhoven,~L.~P.; Petta,~J.~R.; Tarucha,~S.; Vandersypen,~L.
  M.~K. \emph{Rev. Mod. Phys.} \textbf{2007}, \emph{79}, 1217--1265\relax
\mciteBstWouldAddEndPuncttrue
\mciteSetBstMidEndSepPunct{\mcitedefaultmidpunct}
{\mcitedefaultendpunct}{\mcitedefaultseppunct}\relax
\EndOfBibitem
\bibitem[Golovach \latin{et~al.}(2006)Golovach, Borhani, and
  Loss]{Golovach2006}
Golovach,~V.~N.; Borhani,~M.; Loss,~D. \emph{Phys. Rev. B} \textbf{2006},
  \emph{74}, 165319\relax
\mciteBstWouldAddEndPuncttrue
\mciteSetBstMidEndSepPunct{\mcitedefaultmidpunct}
{\mcitedefaultendpunct}{\mcitedefaultseppunct}\relax
\EndOfBibitem
\bibitem[Watzinger \latin{et~al.}(2016)Watzinger, Kloeffel, Vuku\v{s}i\'c,
  Rossell, Sessi, Kuku\v{c}ka, Kirchschlager, Lausecker, Truhlar, Glaser,
  Rastelli, Fuhrer, Loss, and Katsaros]{Watzinger2016}
Watzinger,~H.; Kloeffel,~C.; Vuku\v{s}i\'c,~L.; Rossell,~M.~D.; Sessi,~V.;
  Kuku\v{c}ka,~J.; Kirchschlager,~R.; Lausecker,~E.; Truhlar,~A.; Glaser,~M.;
  Rastelli,~A.; Fuhrer,~A.; Loss,~D.; Katsaros,~G. \emph{Nano Lett.}
  \textbf{2016}, \emph{16}, 6879--6885\relax
\mciteBstWouldAddEndPuncttrue
\mciteSetBstMidEndSepPunct{\mcitedefaultmidpunct}
{\mcitedefaultendpunct}{\mcitedefaultseppunct}\relax
\EndOfBibitem
\bibitem[Kawakami \latin{et~al.}(2014)Kawakami, Scarlino, Ward, Braakman,
  Savage, Lagally, Friesen, Coppersmith, Eriksson, and
  Vandersypen]{Kawakami2014}
Kawakami,~E.; Scarlino,~P.; Ward,~D.~R.; Braakman,~F.~R.; Savage,~D.~E.;
  Lagally,~M.~G.; Friesen,~M.; Coppersmith,~S.~N.; Eriksson,~M.~A.;
  Vandersypen,~L. M.~K. \emph{Nature Nanotechnology} \textbf{2014}, \emph{9},
  666--670\relax
\mciteBstWouldAddEndPuncttrue
\mciteSetBstMidEndSepPunct{\mcitedefaultmidpunct}
{\mcitedefaultendpunct}{\mcitedefaultseppunct}\relax
\EndOfBibitem
\bibitem[Fischer \latin{et~al.}(2008)Fischer, Coish, Bulaev, and
  Loss]{Fischer2008}
Fischer,~J.; Coish,~W.~A.; Bulaev,~D.~V.; Loss,~D. \emph{Phys. Rev. B}
  \textbf{2008}, \emph{78}, 155329\relax
\mciteBstWouldAddEndPuncttrue
\mciteSetBstMidEndSepPunct{\mcitedefaultmidpunct}
{\mcitedefaultendpunct}{\mcitedefaultseppunct}\relax
\EndOfBibitem
\bibitem[Prechtel \latin{et~al.}(2016)Prechtel, Kuhlmann, Houel, Ludwig,
  Valentin, Wieck, and Warburton]{Prechtel2016}
Prechtel,~J.~H.; Kuhlmann,~A.~V.; Houel,~J.; Ludwig,~A.; Valentin,~S.~R.;
  Wieck,~A.~D.; Warburton,~R.~J. \emph{Nature Mat.} \textbf{2016}, \emph{15},
  981--986\relax
\mciteBstWouldAddEndPuncttrue
\mciteSetBstMidEndSepPunct{\mcitedefaultmidpunct}
{\mcitedefaultendpunct}{\mcitedefaultseppunct}\relax
\EndOfBibitem
\bibitem[Ares \latin{et~al.}(2013)Ares, Katsaros, Golovach, Zhang, Prager,
  Glazman, Schmidt, and De~Franceschi]{Ares2013}
Ares,~N.; Katsaros,~G.; Golovach,~V.~N.; Zhang,~J.~J.; Prager,~A.;
  Glazman,~L.~I.; Schmidt,~O.~G.; De~Franceschi,~S. \emph{Appl. Phys. Lett.}
  \textbf{2013}, \emph{103}, 263113\relax
\mciteBstWouldAddEndPuncttrue
\mciteSetBstMidEndSepPunct{\mcitedefaultmidpunct}
{\mcitedefaultendpunct}{\mcitedefaultseppunct}\relax
\EndOfBibitem
\bibitem[van~den Berg \latin{et~al.}(2013)van~den Berg, Nadj-Perge, Pribiag,
  Plissard, Bakkers, Frolov, and Kouwenhoven]{Berg2013}
van~den Berg,~J. W.~G.; Nadj-Perge,~S.; Pribiag,~V.~S.; Plissard,~S.~R.;
  Bakkers,~E. P. A.~M.; Frolov,~S.~M.; Kouwenhoven,~L.~P. \emph{Phys. Rev.
  Lett.} \textbf{2013}, \emph{1110}, 066806\relax
\mciteBstWouldAddEndPuncttrue
\mciteSetBstMidEndSepPunct{\mcitedefaultmidpunct}
{\mcitedefaultendpunct}{\mcitedefaultseppunct}\relax
\EndOfBibitem
\bibitem[Nigg \latin{et~al.}(2017)Nigg, Fuhrer, and Loss]{Nigg2017}
Nigg,~S.~E.; Fuhrer,~A.; Loss,~D. \emph{Phys. Rev. Lett.} \textbf{2017},
  \emph{118}, 147701\relax
\mciteBstWouldAddEndPuncttrue
\mciteSetBstMidEndSepPunct{\mcitedefaultmidpunct}
{\mcitedefaultendpunct}{\mcitedefaultseppunct}\relax
\EndOfBibitem
\bibitem[Kloeffel \latin{et~al.}(2013)Kloeffel, Trif, Stano, and
  Loss]{Kloeffel:PRB2013}
Kloeffel,~C.; Trif,~M.; Stano,~P.; Loss,~D. \emph{Phys. Rev. B} \textbf{2013},
  \emph{88}, 241405(R)\relax
\mciteBstWouldAddEndPuncttrue
\mciteSetBstMidEndSepPunct{\mcitedefaultmidpunct}
{\mcitedefaultendpunct}{\mcitedefaultseppunct}\relax
\EndOfBibitem
\end{mcitethebibliography}

\end{document}